\documentclass[10pt,twoside,twocolumn]{IEEEtran}  
\usepackage[cmex10]{amsmath}
\usepackage{graphicx}
\usepackage{amssymb}
\usepackage[latin1]{inputenc}
\usepackage{color}
\usepackage{cite}
\usepackage{array}
\usepackage{bm}
\usepackage{mathdots}
\usepackage{xcolor}
\usepackage[nonumberlist]{glossaries}
\usepackage{siunitx} 
\usepackage[font=footnotesize]{caption}
\usepackage{subcaption}

\newcommand*{\colorrev}{\color{black}}

\ifCLASSINFOpdf
\else
\fi
\usepackage{hyperref}

\newacronym{3GPP}{3GPP}{3rd generation partnership project}
\newacronym{AN}{AN}{access node}
\newacronym{BD}{BD}{block diagonalization}
\newacronym{BF}{BF}{beamforming}
\newacronym{CDF}{CDF}{cumulative distribution function}
\newacronym{CSI}{CSI}{channel state information}
\newacronym{CSIT}{CSIT}{\gls{CSI} at transmitter}
\newacronym{CV}{CV}{constant velocity}
\newacronym{DL}{DL}{downlink}
\newacronym{OFDM}{OFDM}{orthogonal frequency division multiplexing}
\newacronym{DoA}{DoA}{direction of arrival}
\newacronym{GNSS}{GNSS}{global navigation satellite system}
\newacronym{IoT}{IoT}{Internet of Things}
\newacronym{ITS}{ITS}{intelligent traffic system}
\newacronym{LTE}{LTE}{long term evolution}
\newacronym{LoS}{LoS}{line of sight}
\newacronym{MF}{MF}{matched filter}
\newacronym{MRC}{MRC}{maximum ratio combining}
\newacronym{MIMO}{MIMO}{multiple-input multiple-output}
\newacronym{mmWave}{mmWave}{millimeter wave}
\newacronym{MU-MIMO}{MU-MIMO}{multi-user \gls{MIMO}}
\newacronym{MU-MISO}{MU-MISO}{multi-user multiple-input single-output}
\newacronym{NLoS}{NLoS}{non line of sight}
\newacronym{OTDoA}{OTDoA}{observed time difference of arrival}
\newacronym{QoE}{QoE}{quality of experience}
\newacronym{RBF}{RBF}{receive \gls{BF}}
\newacronym{REM}{REM}{radio environment map}
\newacronym{RE}{RE}{resource element}
\newacronym{QPSK}{QPSK}{quadrature phase-shift keying}
\newacronym{RF}{RF}{radio frequency}
\newacronym{RS}{RS}{reference signal}
\newacronym{RSSI}{RSSI}{received signal strength indication}
\newacronym{RRM}{RRM}{radio resource management}
\newacronym{SINR}{SINR}{signal-to-interference-plus-noise ratio}
\newacronym{SLAM}{SLAM}{simultaneous localization and mapping}
\newacronym{TD}{TD}{time domain}
\newacronym{TDD}{TDD}{time division duplex}
\newacronym{ToA}{ToA}{time of arrival}
\newacronym{TTI}{TTI}{transmission time interval}
\newacronym{UDN}{UDN}{ultra-dense network}
\newacronym{UKF}{UKF}{unscented Kalman filter}
\newacronym{UL}{UL}{uplink}
\newacronym{UE}{UE}{user equipment}
\newacronym{WLAN}{WLAN}{wireless local area network}
\newacronym{ZF}{ZF}{zero forcing}
\newacronym{ID}{ID}{identifier}
\newacronym{EKF}{EKF}{extended Kalman filter}
\newacronym{AoA}{AoA}{angle-of-arrival}
\newacronym{DCI}{DCI}{DL control information}

\begin{document}
%
\title{High-Efficiency Device Positioning and Location-Aware Communications in Dense 5G Networks}
%
%
%

\author{Mike Koivisto, \IEEEmembership{Student Member, IEEE}, Aki Hakkarainen, M\'ario Costa, \IEEEmembership{Member, IEEE},\\  Petteri Kela, \IEEEmembership{Member, IEEE}, Kari Lepp\"anen, and Mikko Valkama, \IEEEmembership{Senior Member, IEEE}\thanks{M. Koivisto, A. Hakkarainen, and M. Valkama are with the Department of Electronics and Communications Engineering, Tampere University of Technology, Tampere 33720, Finland (e-mail: mike.koivisto@tut.fi; aki.hakkarainen@tut.fi; mikko.e.valkama@tut.fi).}\thanks{M. Costa, P. Kela and K. Lepp\"anen are with Huawei Technologies Oy (Finland), Ltd., Helsinki 00180, Finland (e-mail: mariocosta@huawei.com; petteri.kela@huawei.com; kari.leppanen@huawei.com). P. Kela is also with the Department of Communications and Networking, Aalto University.}\thanks{Multimedia material available at {\color{blue} \url{http://www.tut.fi/5G/COMMAG16/}}}
\thanks{This work has been accepted for publication in IEEE Communications Magazine. This is the revised version of the original work and it is currently in press. Copyright may be transferred without notice, after which this version may no longer be accessible.}}

\maketitle

\begin{abstract}

In this article, the prospects and enabling technologies for high-efficiency device positioning and location-aware communications in emerging 5G networks are reviewed. We will first describe some key technical enablers and demonstrate {\colorrev by means of realistic} ray-tracing and map based evaluations that positioning accuracies below one meter can be achieved by properly fusing direction and delay related measurements on the network side, even when tracking moving devices. We will then discuss the possibilities and opportunities that such high-efficiency positioning capabilities can offer, not only for location-based services in general, but also for the radio access network itself. In particular, we will demonstrate that geometric location-based beamforming schemes become technically feasible, which can offer substantially reduced reference symbol overhead compared to classical full \gls{CSI}-based beamforming. At the same time, substantial power savings can be realized in future wideband 5G networks where acquiring full \gls{CSI} calls for wideband reference signals while location estimation and tracking can, in turn, be accomplished with narrowband pilots.
 
\end{abstract}


%
\IEEEpeerreviewmaketitle

\section{Introduction} \label{sec:introduction}

\begin{figure}[!t]
    \centering
    \begin{subfigure}{\columnwidth}
        \centering
        \includegraphics[width=8.85cm]{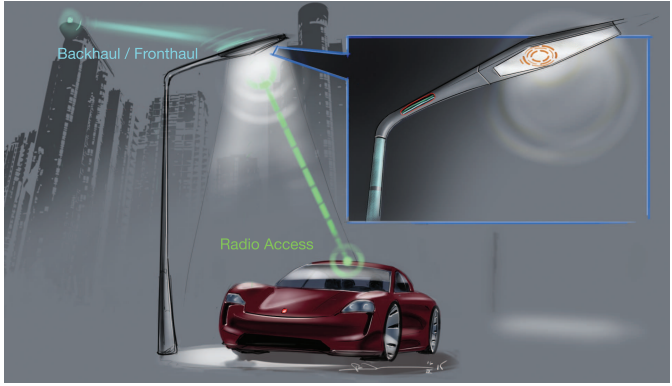}
        \caption{}
        \label{fig_lamp_post}
    \end{subfigure}  
    \\[12pt]
    \begin{subfigure}{\columnwidth}
        \centering
        \includegraphics[width=8.85cm]{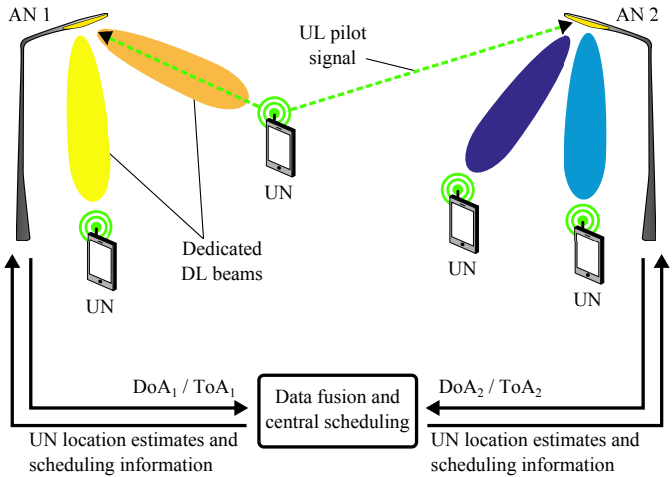}
        \caption{}
        \label{fig_overall}
    \end{subfigure}
    \caption{Illustration of a 5G network where (a) AN, deployed in a lamp post, provides a LoS connection to a nearby UE and (b) ANs estimate DoAs/ToAs of the UEs based on UL pilot signals. The obtained estimates are then communicated to a fusion center providing the final location estimate which, in turn, enables geometric DL beamforming.}
    \label{fig_ANs}
    \vspace{-6pt}
\end{figure}

\IEEEPARstart{F}{uture} 5G networks are expected to provide huge improvements in the capacity, number of connected devices, energy efficiency, and latencies when compared to the existing communications systems~\cite{Osseiran14, huawei_5g_air_interface}. These features will be enabled by the combination of higher bandwidths, advanced antenna technologies, and flexible radio access solutions, among others. Especially in urban environments, 5G networks are also expected to consist of densely distributed \glspl{AN}~\cite{huawei_5g_air_interface} 
located, e.g., in lamp posts above the streets as illustrated in Fig.~\ref{fig_lamp_post}. Consequently, a single \gls{UE} in such dense networks {\colorrev is within coverage range to} multiple closely located \glspl{AN} at a time. Such short \gls{UE}-\gls{AN} distances
provide obvious benefits for communications, e.g., due to lower propagation losses and shorter propagation times, but interestingly can also enable highly accurate \gls{UE} positioning. Altogether, 5G networks allow for many opportunities regarding acquisition and exploitation of \gls{UE} location information in unforeseen manners~{\colorrev\cite{Werner15, koivisto_joint_2016}}. This is the leading theme of this article.

\urldef\gforum\url{http://kani.or.kr/5g/whitepaper/2015%205G_Forum_White_Paper_Service.pdf}

One of the improvements in 5G networks concerns the positioning accuracy.
It is stated, e.g., in~\cite{ngmn_5g, 5g-ppp_vision_2015, 5g_forum_5g_2015}, {\colorrev and~\cite{3GPP_TS_38_913}\footnote{See also 3GPP technical report 22.862, v.14.1.0.}}, that 5G should provide a positioning accuracy in the order of one meter or even below. That is significantly better than the accuracy of a couple of tens of meters provided in \gls{LTE} systems by \gls{OTDoA}-based techniques. The required positioning accuracy in 5G networks will outperform also commercial \glspl{GNSS} where the accuracy is around \SI{5}{m}, and \gls{WLAN} fingerprinting resulting in a \mbox{\SIrange[range-phrase = --]{3}{4}{m}} accuracy. 
Another improvement that 5G networks may provide concerns the energy efficiency of positioning. This stems from the common assumption that 5G networks will exploit frequently transmitted \gls{UL} pilot signals for channel estimation purposes at the {\colorrev \glspl{AN}}. These signals can be used also for positioning in a network-centric manner where the \gls{UE} location is estimated either {\colorrev independently} in the \glspl{AN} or in a centralized fusion center{\colorrev, assuming known AN locations,} and thus no calculations are needed in the mobile \glspl{UE}. Note that this is a considerable difference to the device-centric positioning, e.g., \gls{GNSS}, where the mobile \glspl{UE} are under heavy computational burden. Therefore, network-centric positioning techniques provide significant power consumption improvements and enable ubiquitous high-accuracy positioning that can run in the background continuously. 
{\colorrev Such a functionality decreases also the signaling overhead when the location information is to be used on the network side, but on the other hand, requires additional care for privacy as the positioning is not carried out at the \glspl{UE} themselves.}
As a third improvement in 5G-based positioning, {\colorrev regardless whether it is network- or device-centric, location information can be obtained} in complete independence of \gls{UE}-satellite connections everywhere under the network coverage area, including also challenging indoor environments.

\urldef\whereone\url{http://www.ict-where.eu/}
\urldef\wheretwo\url{http://www.ict-where2.eu/}

The aim of this article is to discuss the technical enablers of envisioned device positioning in 5G networks, and to promote the prospects of the obtained location-awareness. In this regard, focus is given to {\colorrev location-based} communication and network management techniques such as location-based beamforming as well as mobility and \gls{RRM}{\colorrev~\cite{SDM14}\footnote{\colorrev{See also the WHERE and WHERE2 projects at \mbox{\whereone} and \mbox{\wheretwo}}}}. We recognize that \gls{UE} location information can be exploited by the \gls{UE} itself as well as shared with third parties, thus allowing for innovative location-based applications to emerge.
Particularly, we will focus on the connected car application, being identified, e.g., in~\cite{ngmn_5g} as one key application and target for future 5G mobile communication networks, with a minimum of $2000$~connected vehicles per km$^2$ and at least \SI{50}{Mbps} in \gls{DL} {\colorrev throughput}.
Now, facilitating such greatly enhanced connected vehicle applications, having a 5G network with built-in capability to localize and track vehicles is a very tempting prospect. Furthermore, location information is a central element towards self-driving cars, \glspl{ITS}, drones as well as other kinds of autonomous vehicles and robots which are envisioned to be part of not only the future factories, but the overall future society within the next $5-10$~years.


%
%
%
%

\section{5G Networks and Positioning Prospects} \label{sec:network_and_positioning}

{\colorrev \subsection{Technical Properties of 5G Radio Networks}}

Generally, it is expected that network densification will play an important role in achieving demanding requirements of 5G networks. The inter-site distance of \glspl{AN} in such \glspl{UDN} is envisioned to range from a few meters up to a few tens of meters, e.g., assuming several \glspl{AN} per room indoors and an \gls{AN} on each lamp post outdoors~\cite{Osseiran14}. Moreover, these 5G \glspl{AN} are expected to be equipped with smart antenna solutions, such as antenna arrays supporting \gls{MIMO} techniques~\cite{5g-ppp_vision_2015}. Such antenna technologies are suitable for effective communications as well as accurate \gls{DoA} estimation, which in turn allows for high-accuracy positioning. Furthermore, it is argued that devices tend to be in \gls{LoS} condition with one or multiple \glspl{AN} due to network densification, which is a favorable condition not only for communications but also for positioning purposes. 

{\colorrev It} is commonly agreed that 5G technologies will require wide bandwidths in order to meet the envisioned capacity requirements. Therefore, 5G networks will most likely operate at higher frequency bands, including \glspl{mmWave}, where the availability of unallocated spectrum is considerably higher. Such high frequency bands together with \glspl{UDN} can provide very high overall system capacity and enable an efficient frequency reuse~\cite{5g-ppp_vision_2015}. However, with the envisioned high frequencies, the propagation conditions become more demanding due to, e.g., larger propagation losses. Hence, the effective range between transmitting and receiving elements is relatively short which also emphasizes the importance of expected \glspl{UDN}. Furthermore, the utilization of effective antenna solutions become more practical as a result of shorter wavelengths, and consequently due to smaller physical size of antenna elements. {\colorrev In addition to the potential frequency bands above \SI{6}{GHz}, also frequencies below \SI{6}{GHz} are expected to be used in 5G networks~\cite{ngmn_5g}.} Apart from a communication perspective, the envisioned wide bandwidths enable also very accurate \gls{ToA} estimates which in turn provide an opportunity for positioning with remarkably high accuracy \cite{Werner15}.

In contrast to the earlier cell-centric architectures, it is currently under discussion whether 5G networks will be developed in a more device-centric manner. Moreover, it is envisioned that 5G networks could also provide improved \gls{QoE} at cell borders with only a minimal system-performance degradation compared to earlier systems~\cite{kela_borderless_2015}. This development enables tailoring of a particular communication session and the functions of the associated \glspl{AN} to a connected device or service instead of obtaining services from the \gls{AN} commanding the specific cell. 
In such device-centric architecture, a given device can periodically send \gls{UL} signals to connected \glspl{AN} in which \gls{UL} reference signals are used for channel estimation, but they can also be employed for network-centric positioning as illustrated in Fig.~\ref{fig_overall}. Furthermore, future 5G networks are expected to operate with relatively short radio frames 
resulting in availability of frequent location information about a transmitting device. 

{\colorrev\subsection{Leveraging Location-Awareness in 5G Networks}}
\label{sec:prospects}

\begin{figure}[!t]
    \centering
    \begin{subfigure}{\columnwidth}
        \centering
        \includegraphics[width=8.85cm]{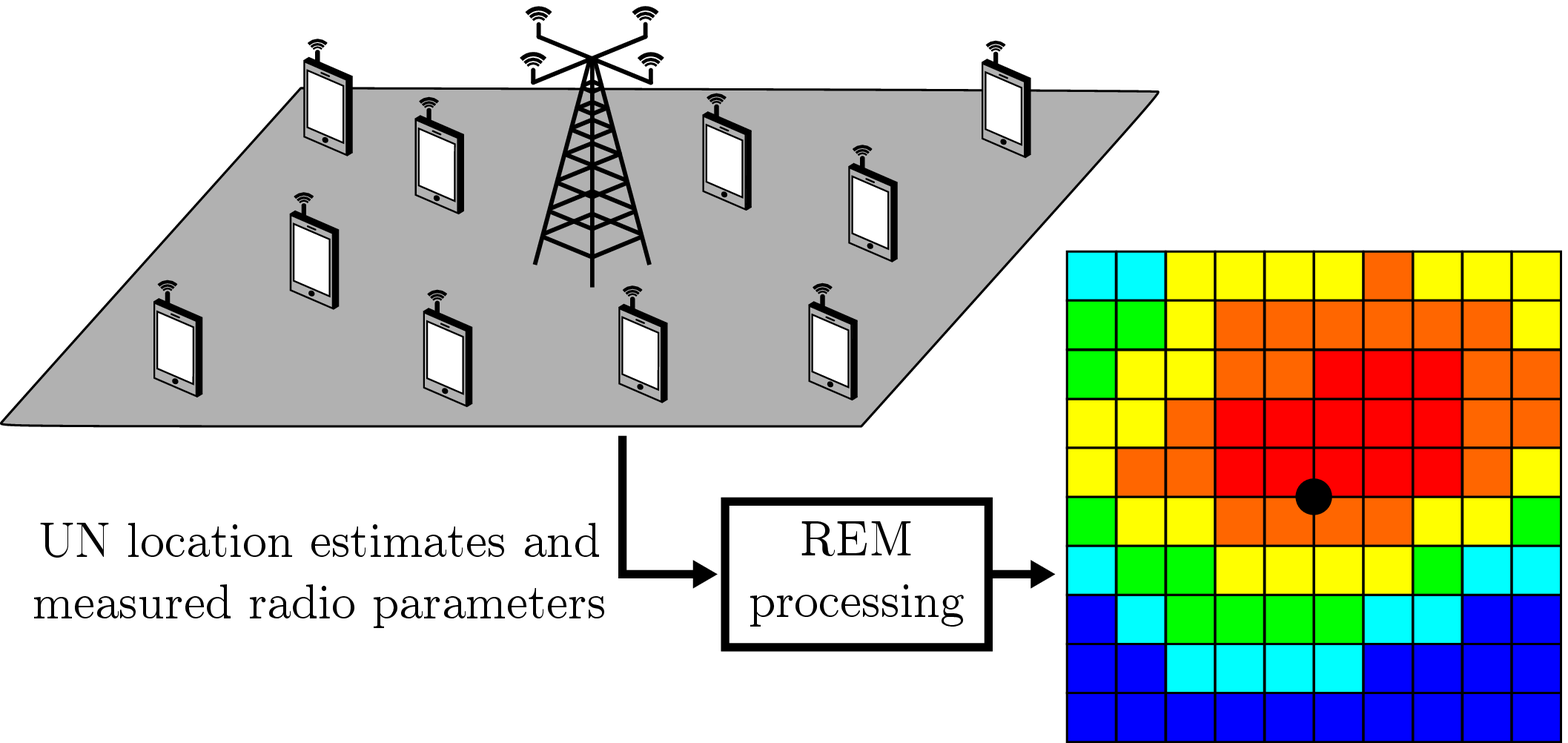}
        \caption{}
        \label{fig_REM}
    \end{subfigure}
    \\[12pt]
    \begin{subfigure}{\columnwidth}
        \centering
        \includegraphics[width=8.85cm]{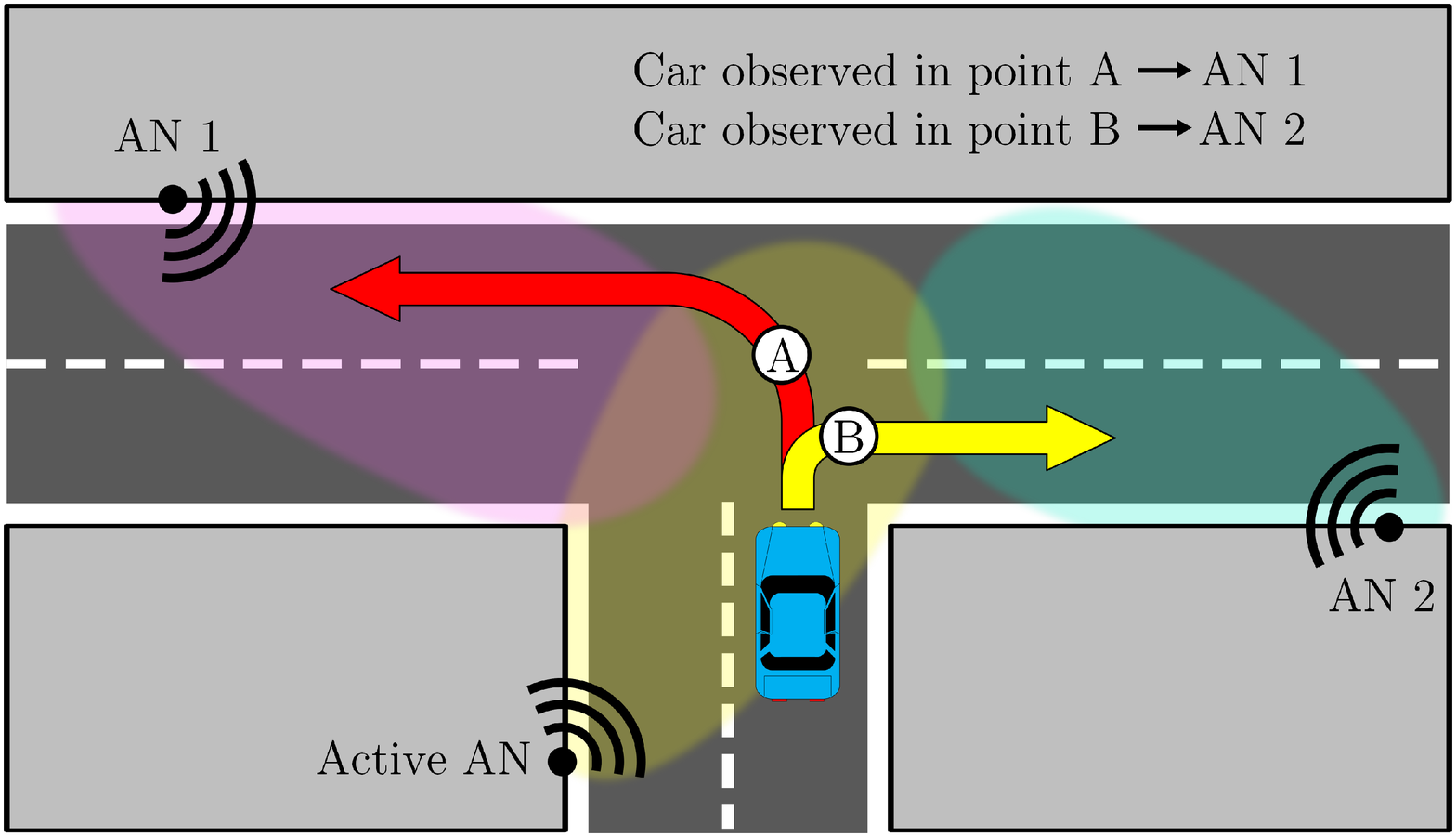}
        \caption{}
        \label{fig_proactive}
    \end{subfigure}  
    \\[12pt]
    \begin{subfigure}{\columnwidth}
        \centering
        \includegraphics[width=6.55cm]{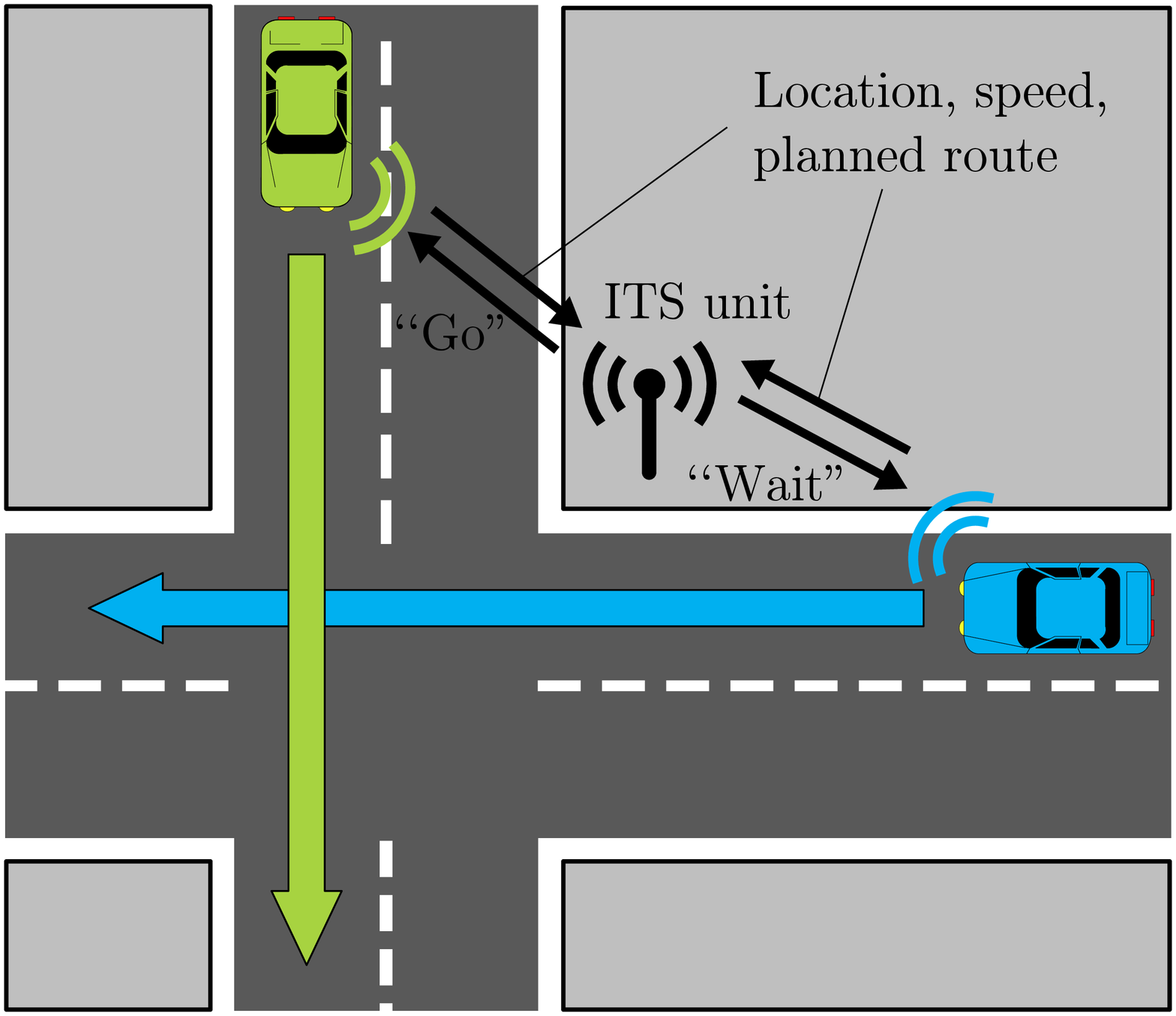}
        \caption{}
        \label{fig_ITS}
    \end{subfigure}  
    \caption{Illustrations of {\colorrev selected} positioning prospects in 5G a) \gls{REM} generation, b) proactive \gls{RRM} for a car whose location is being tracked, and c) \gls{ITS}-based traffic control and collision avoidance with self-driving cars.}
    \label{fig_prospects}
    \vspace{-6pt}
\end{figure}

Continuous positioning provides awareness not only of the current but also of the past \gls{UE} locations and thus the network is able to carry out \gls{UE} tracking. When the \gls{UE} location and movement information is processed by predictive algorithms, the network can even predict the \gls{UE} locations to some extent. Availability of such location information in turn enables a wide selection of entirely new features and services in 5G networks. First, location-awareness can be used for communications purposes by enhancing the utilization of spatial dimension, e.g., by geometric beamforming~\cite{kela_location_based_beamforming_2016} and sophisticated spatial {\colorrev interference mitigation}. These features allow for multiplexing a high density of \glspl{UE} and provide significant throughput improvements for high-mobility \glspl{UE}, as illustrated in Section~\ref{sec:beamforming_and_mobility}. Second, a combination of location information and measured radio parameters over a long time period {\colorrev enables} the construction of \glspl{REM}, depicted in Fig.~\ref{fig_REM}, which, in turn, can open many opportunities in terms of proactive \gls{RRM}{\colorrev~\cite{SDM14}}. Particularly, knowledge of large-scale fading and location-based radio conditions can be utilized for {\colorrev\it\gls{RRM} purposes} without the need of knowing the instantaneous channel information between the \gls{AN} and \gls{UE}. Therefore, the network is able to carry out proactive allocation of active \glspl{UE} to nearby \glspl{AN} such that, e.g., power consumption, load balancing and latencies are optimized as depicted in Fig.~\ref{fig_proactive}. Location-awareness can improve network functionalities also 
by enabling proactive location-based backhaul routing such that the \gls{UE}-specific data can be communicated with a high robustness and low end-to-end latency. 

{\colorrev The obtained location-awareness can be exploited also in the \glspl{UE} as well as by third parties for providing other than purely communications type of services. Taking traffic and cars as an example, up-to-date location information and predicted \gls{UE} trajectories can provide remarkable improvements, e.g., in terms of traffic flow, safety and energy efficiency. When comprehensively gathered car location information is shared with \glspl{ITS}, functionalities such as traffic monitoring and control can be enhanced. Accurate location information is needed also in the cars themselves, e.g., for navigation purposes, especially when considering autonomous and self-driving cars. Location-awareness is required also for collision avoidance. Within communications range cars can report their location directly to other cars, but when the link between the cars is blocked, location notifications are transmitted in collaboration with \glspl{ITS} as illustrated in Fig.~\ref{fig_ITS}. Naturally, the demands and functionalities regarding self-driving cars cannot be met everywhere and at all times by existing communications systems and satellite-based positioning. Consequently, advanced communications capabilities and network-based positioning in 5G is likely to play an important role in the development of self-driving car systems}.\\[2pt]

\section{Enabling Technologies for High-Efficiency Network-centric Positioning} \label{sec:technologies}

\subsection{State-of-the-Art}
\label{sec:state}

Dense networks are characterized by radio channels that are dominated by the \gls{LoS}-path. 
For example, the typical Rice-factor{\colorrev, being a power ratio between the LoS component and all other propagation paths,} in urban micro-cell environments is around \SI{10}{\decibel}, even in sub-\SI{6}{\giga \hertz}~\cite{metis_channels}. Additionally, network densification  increases the \gls{LoS} probability between \glspl{UE} and \glspl{AN}. As an example, 3GPP employs a channel model based on extensive measurements in which the \gls{LoS} probability is higher than \SI{0.7}{} for a maximum \gls{UE}-\gls{AN} distance of \SI{35}{m}. 

Determining the \glspl{AN} that are in \gls{LoS} condition to a given \gls{UE} is important since it allows estimating and tracking the directional parameters of the \gls{LoS}-path, in addition to the time-of-flight and clock-offsets, thus greatly improving the \gls{UE} positioning accuracy. Particularly, the \gls{LoS} condition of a radio link may be assessed by estimating the corresponding Rice-factor. A multichannel observation is obtained for each \gls{UL} reference signal given that a multicarrier waveform and multiantenna \glspl{AN} are employed. Sequential estimation of the Rice-factor can be accomplished, e.g., with a particle filter due to the non-Gaussian nature of the amplitude distribution of the \gls{UL} multicarrier channel. Finally, \gls{LoS} detection can be accomplished using a likelihood-ratio test, or a model selection technique. In case all \glspl{AN} are in \gls{NLoS} to the \gls{UE}, {\colorrev coarse} network-centric positioning can still be achieved using radio frequency fingerprinting, received signal strength indicator and cell-identifier, among others \cite{SDM14}.

Multicarrier waveforms offer a versatile approach for estimating ranges between a given \gls{UE} and multiple \glspl{AN}~\cite{SDM14}. Relying solely on \gls{UL} reference signals makes it possible to synchronize the \glspl{AN} as well as the \gls{UE}, in addition to estimating the \glspl{ToA} of the \gls{LoS}-paths~\cite{Werner15, koivisto_joint_2016}. 
{\colorrev
The actual sequential estimation of the \glspl{ToA} and clock-offsets can be accomplished with different Bayesian filters either in a
cascaded or fully centralized manner depending on the network architecture, baseband processing capabilities of the \glspl{AN}, and backhaul capacity. 
Note that the UL reference signals can also provide additional information for \gls{UE} positioning when utilized, e.g., for tracking Doppler-shifts.
}

\glspl{AN} with multiantenna transceivers allow for estimating the \gls{DoA} of the \gls{LoS}-path from \gls{UL} reference signals, and such an information can be used for \gls{UE} positioning. Planar or conformal arrays, such as circular or cylindrical antenna arrays, make it possible for estimating elevation and azimuth arrival angles, and enable 3D positioning. Bayesian filtering techniques can also be employed for tracking the \glspl{DoA} of the \gls{LoS}-paths from mobile \glspl{UE} as well as fusing the \glspl{ToA} and \glspl{DoA} in order to allow for joint \gls{UE} positioning and network synchronization~\cite{Werner15, koivisto_joint_2016}. \glspl{AN} with analog beamforming structures and sectorized antennas can also be exploited for \gls{UE} positioning and tracking~\cite{WWHCV15}.


{\colorrev Due to the non-linear nature of the involved state-space models, estimation and tracking can be carried out with different non-linear Bayesian filtering techniques. In this article, the tracking processes are carried out using the \gls{EKF} due to its highly accurate estimation performance and low computational complexity compared to, e.g., particle filters and the \gls{UKF}. In general, within the \gls{EKF}, the state of a system is first propagated through a linearized state evolution model and this prediction is, thereafter, updated using the obtained measurements and a linearized measurement model, through which the state is associated with the measurements~\cite{koivisto_joint_2016}.}

Finally, the techniques overviewed in this section for \gls{UE} positioning can also be employed for estimating the locations of the \glspl{AN}. For example, a few well-surveyed \glspl{AN} can be used for finding the locations of neighboring \glspl{AN}, which in turn may be used as new anchors. Such a procedure is useful since surveying all \glspl{AN} would increase the deployment cost of \glspl{UDN} significantly. {\colorrev Alternatively, joint \gls{UE} tracking and \glspl{AN} positioning can be achieved using techniques stemming from \gls{SLAM} \cite{bruno_wislam_2011}. These techniques are versatile but the cost is an increase in computational complexity due to the large number of parameters to be estimated.}

\subsection{Tracking of Directional Parameters using EKFs}
\label{sec:DoA_results}

We start by {\colorrev demonstrating} the performance of \glspl{EKF} in tracking the directional parameters of the \gls{LoS}-path. We consider the case where both the \glspl{AN} and \glspl{UE} are equipped with multiantenna transceivers. Two schemes are considered, namely a network-centric approach and a decentralized scheme. {\colorrev In} the network-centric approach, the arrival and departure angles of the \gls{LoS}-path between a \gls{UE} and an \gls{AN} are tracked jointly at the \gls{AN}\footnote{The departure angles can only be retrieved from the arrival angles if the orientation of the \gls{UE}'s array is known. The network-centric approach requires that the calibration data of the \gls{UE}'s array is acquired by the \gls{AN}, e.g., over a \gls{UL} control channel.}. The \gls{UE} transmits periodically \gls{UL} reference signals from all of its antenna elements. Each \gls{UE} antenna element is assigned a single subcarrier, which is different from {\colorrev those} used by the other antennas. 
The departure angles are transmitted to the \gls{UE} on a \gls{DL} control channel.

The decentralized scheme consists in tracking the double-directional parameters of the \gls{LoS}-path independently at the \gls{AN} and \gls{UE}. Such a scheme is based on narrowband \gls{UL} transmissions from a single {\colorrev antenna element of a \gls{UE}}. This allows the \gls{AN} to track the arrival angles of the \gls{LoS}-path. These arrival angles are used for designing a beamforming weight-vector that is exploited by the \gls{AN} to transmit a beamformed \gls{DL} reference signal towards the \gls{UE}. This makes it possible for the \gls{UE} to track the arrival angles, and thus design the receive beamforming weight-vector. The transmit and receive beamforming weight-vectors designed in this fashion are compared to \gls{CSIT}-based precoding schemes in Section \ref{sec:beamforming_and_mobility}.


\begin{figure}[!t]
    \centering
    \includegraphics[scale=0.52]{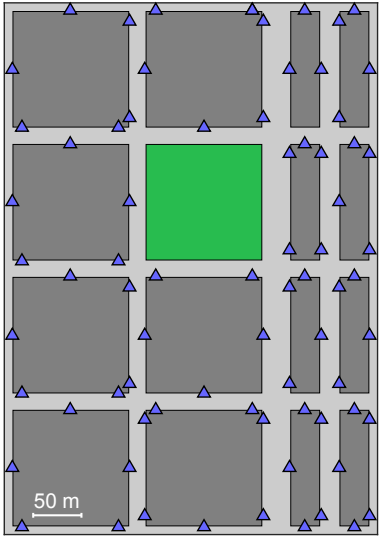}
    \caption{METIS Madrid grid layout, from~\cite{metis_channels}, where ANs (blue triangles) are distributed densely along the streets.}
    \label{fig_madrid}
    \vspace{-6pt}    
\end{figure}

\begin{figure}[!t]
    \centering
    \includegraphics[width=\columnwidth,trim={0.8cm 0.5cm 1.2cm 0.5cm},clip]{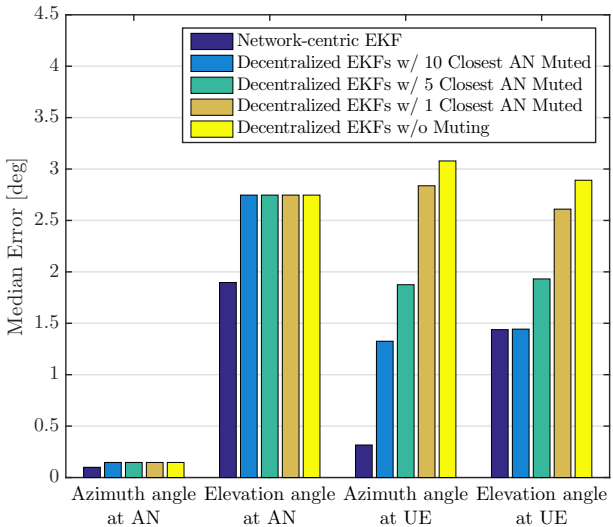}
    \caption{Accuracy of tracking the arrival and departure angles with \glspl{EKF} in terms of the median error. In network-centric \gls{EKF}, the \gls{UE} transmits \gls{UL} reference signals from all antenna elements and the \gls{AN} tracks both arrival and departure angles of the \gls{LoS}-path. In decentralized \gls{EKF}, the \gls{UE} transmits \gls{UL} reference signals from a single antenna element which is used by the \gls{AN} to track the arrival angles of the \gls{LoS}-path with an \gls{EKF}. Such directional parameters are employed in order to design a \gls{DL} beamformed reference signal that {\colorrev then} allows the \gls{UE} to track and design a similar receive beamforming vector.}
    \label{fig_angle_results}
    \vspace{-6pt}    
\end{figure}

The performance of both network-centric and decentralized approaches have been analyzed with a \gls{TDD} based 5G simulator. In particular, we have considered a \gls{UDN} composed of $74$ \glspl{AN} with a deployment identical to that illustrated in Fig.~\ref{fig_madrid}. The \glspl{AN} are equipped with circular arrays composed of $20$ dual-polarized 3GPP patch elements \SI{5}{m} above the ground. 
The \glspl{UE} are equipped with circular arrays composed of $4$ dual-polarized elements (cross-dipoles). The transmit power budget of the \glspl{UE} is \SI{10}{dBm} while that of the \glspl{AN} is \SI{23}{dBm}. The \glspl{UE} take different routes {\colorrev through} the Madrid grid (see Fig.~\ref{fig_madrid}) with velocities of \SIrange{30}{50}{km/h}. The carrier-frequency is \SI{3.5}{GHz} and the METIS map-based ray tracing channel model \cite{metis_channels} has been employed. An \gls{OFDM} waveform is used in both \gls{UL} and \gls{DL}. The subcarrier spacing is \SI{240}{\kilo\hertz} and the \gls{TTI} equals \SI{200}{\micro\second}. The \gls{UL} and \gls{DL} reference signals are Zadoff-Chu sequences, similar to those used in \gls{LTE}. The pilots employed by the network-centric and decentralized \glspl{EKF} for tracking the double-directional parameters are transmitted on a single subcarrier in a frequency-hopping manner spanning \SI{10}{\mega\hertz}. Such \gls{UL} and \gls{DL} pilots are transmitted on every $500^\text{th}$ \gls{TTI}. Hence, the \gls{UL} and \gls{DL} beaconing rate is \SI{10}{beacons/s}. The latency between \gls{UL} and \gls{DL} pilots in the decentralized scheme is $2$ \glspl{TTI}.

Fig.~\ref{fig_angle_results} illustrates the performance of both network-centric and decentralized \glspl{EKF} in tracking the double-directional parameters of the \gls{LoS}-path in terms of the median error. In the network-centric \gls{EKF}, the \gls{UL} beacons received at the \glspl{AN} are impaired by uncoordinated interference due to \glspl{UE} transmitting simultaneously roughly \SI{250}{\meter} away from the receiving \glspl{AN}. {\colorrev The performance difference in azimuth-angle estimation at the \gls{AN} and \gls{UE} is due to the larger array aperture of the former. However, the elevation angle estimates at the \glspl{UE} outperform those obtained at the \glspl{AN}. This is explained by the highly directive beampatterns employed at the \glspl{AN} which decrease the effective aperture of the \glspl{AN}' arrays in the elevation domain. In particular, the beampatterns of the 3GPP patch elements composing the arrays at the \glspl{AN} are characterized by a large attenuation at the poles, thus decreasing the estimation accuracy that can be obtained for the elevation angles.}

In the decentralized \gls{EKF}, the \glspl{AN} transmit \gls{DL} reference signals to \SI{8}{} \glspl{UE} simultaneously (\SI{20}{\percent} of the spatial degrees-of-freedom). Such \gls{DL} reference signals are impaired by similar pilots transmitted by neighboring \glspl{AN} (unless {\colorrev being} muted). Results {\colorrev in Fig.~\ref{fig_angle_results}} show that muting neighboring \glspl{AN} leads to improved performance on the azimuth and elevation angles at the \glspl{UE} due to reduced \gls{DL} interference. Such an interference coordination does not influence the performance of the estimated azimuth and elevation angles at the \glspl{AN} since these parameters are {\colorrev estimated} from \gls{UL} reference signals. The network-centric EKF outperforms the decentralized EKF since all parameters are estimated and tracked jointly. The cost is an increase of the computational complexity and {\colorrev required} control channel capacity.

\subsection{Positioning Accuracy using Cascaded EKFs}
\label{sec:positioning_results}

Next we assume that the \glspl{DoA} and \glspl{ToA} are acquired using the network-centric \gls{EKF}-based approach deployed at \glspl{AN} as described in Section~\ref{sec:DoA_results} {\colorrev with more details available in~\cite{koivisto_joint_2016}}. These spatial and temporal estimates from all the \gls{LoS}-\glspl{AN} can be thereafter fused into 3D \gls{UE} location estimates using an additional positioning and synchronization \gls{EKF}, thus assembling a cascaded \gls{EKF} solution within a network {\colorrev as a whole~\cite{Werner15, koivisto_joint_2016}}. In addition to 3D location estimates, the latter \gls{EKF} can be simultaneously used for tracking the valuable clock offset estimates of unsynchronized \glspl{UE} and \gls{LoS}-\glspl{AN}. In order to demonstrate the performance of the cascaded \gls{EKF}, two alternative scenarios for synchronization are considered. In the first scenario, the \glspl{UE} have unsynchronized clocks with drifts whereas the \glspl{AN} are assumed to be synchronized among each other. In the second scenario, the ANs have also mutual clock-offsets, which are not fundamentally varying over time, whereas the clocks within the \glspl{UE} are again drifting as mentioned above. Such scenarios are later denoted as Pos\&Clock \gls{EKF} and Pos\&Sync \gls{EKF}, respectively~\cite{koivisto_joint_2016}.

\begin{figure}[!t]
    \centering
    \includegraphics[width=\columnwidth,trim={0.8cm 0.5cm 1.2cm 0.5cm},clip]{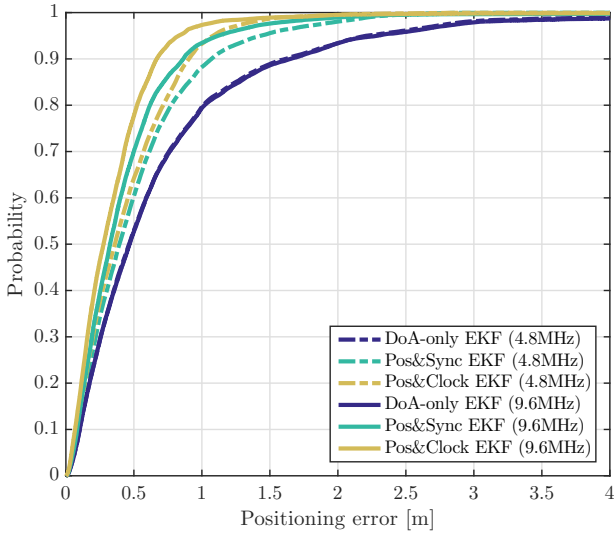}
    \caption{{\colorrev\Glspl{CDF} for 2D positioning errors with \SI{4.8}{MHz} and \SI{9.6}{MHz} \gls{RS} bandwidths over random routes through the Madrid map. Pos\&Clock EKF refers to synchronized \glspl{AN} whereas Pos\&Sync EKF refers to unsynchronized network elements.}}
    \label{fig_pos_results}
    \vspace{-6pt}    
\end{figure}

Considering the radio interface numerology described in Section~\ref{sec:DoA_results} and exploiting a \gls{CV} motion model for the \glspl{UE} attached to vehicles with a maximum speed of \SI{50}{km/h}, the performance of the Pos\&Clock and Pos\&Sync \glspl{EKF} are compared with the classical \gls{DoA}-only \gls{EKF} using both \SI{4.8}{MHz} and \SI{9.6}{MHz} \gls{RS} bandwidths. Since only automotive applications are considered here, more appealing 2D positioning approach was used in the evaluations. {\colorrev The 2D positioning results in terms of \glspl{CDF} are depicted in Fig.~\ref{fig_pos_results} after averaging over multiple random trajectories on the Madrid grid.} Based on the results, the cascaded \glspl{EKF} can provide extremely accurate location estimates for the \glspl{UE} even in the case of unsynchronized \glspl{AN}. As expected, Pos\&Clock and Pos\&Sync \glspl{EKF} outperform the \gls{DoA}-only \gls{EKF} due to the additional \gls{ToA} estimates. Because of a better time resolution, \SI{9.6}{MHz} \gls{RS} bandwidth implies more accurate {\colorrev\gls{ToA} estimates}, and consequently, more accurate positioning can be obtained. {\colorrev Despite the fact that the Pos\&Clock \gls{EKF} is more accurate than the Pos\&Sync EKF due to the synchronized \glspl{AN}, both methods can achieve the envisioned sub-meter positioning accuracy of future 5G networks~\cite{5g-ppp_vision_2015, 5g_forum_5g_2015} with a probability of at least 93\% when using the \SI{9.6}{MHz} bandwidth. In addition to high-accuracy positioning performance, both Pos\&Clock and Pos\&Sync EKFs are able to track also the clock offsets of the \glspl{UE} and \glspl{AN} with an extremely high accuracy.
Finally, due to being able to estimate both azimuth and elevation DoAs in addition to ToAs, the positioning EKF can also facilitate 3D and single AN based positioning\footnote{\colorrev Video of 3D and single AN-based positioning is available at \url{http://www.tut.fi/5G/COMMAG16}.}.}



\section{Location-Based Geometric Beamforming and Mobility Management} \label{sec:beamforming_and_mobility}

Network densification and accurate \gls{UE} positioning in 5G will open new opportunities also for \gls{RRM} and \gls{MIMO}. Especially \gls{MU-MIMO} is seen as a promising solution for 5G as it  
enables \gls{MIMO} gains also with simple single antenna \glspl{UE}. 
As discussed in Section~\ref{sec:state}, \glspl{UE} in \glspl{UDN} are close to an \gls{AN} with a high \gls{LoS} probability. This makes it possible to design and adopt geometric beams at transmitters without the need to estimate the full-band \gls{CSIT}{\colorrev~\cite{SDM14, kela_borderless_2015, kela_location_based_beamforming_2016}}. This is enabled by using the estimated elevation and azimuth angles relative to the \gls{AN}'s coordinate system. The synthesized \gls{MU-MISO} matrix can then be formed comprising only \gls{LoS}-paths for all served \glspl{UE}. One significant benefit of such beamforming scheme is that full-band \gls{UL} reference signals, traditionally employed for obtaining \gls{CSIT}, can be replaced with narrowband \gls{UL} pilots. 
This will allow for substantial energy savings, especially on \gls{UE} side, which is a very important aspect in future wideband 5G networks. In addition to transmit beamforming, the location-based approach can be used also for calculating the receive filters at \glspl{UE} when high-accuracy \gls{DoA} estimates of the desired signals are available.

In addition to \gls{MU-MIMO} beamforming, accurate positioning is also a key enabler for paradigm shift from classical cellular networks towards device-centric borderless networks with centralized control entities. When {\colorrev the} network is constantly keeping track of \gls{UE} locations, it can assign a set of serving \glspl{AN} for each \gls{UE}. Then data for each \gls{UE} is partially {\colorrev or} fully available at some small set of nearby \glspl{AN} as also outlined in \cite{huawei_5g_air_interface}. This enables ultra-short latencies and borderless \gls{QoE} with seamless mobility decreasing handover latencies~\cite{kela_borderless_2015}. Furthermore, such device-centric mobility approach can reduce the energy and resource consuming cell measurement and reporting burden of legacy cellular systems.


\subsection{Evaluation Setup}

We consider a similar setup as in Sections~\ref{sec:DoA_results} and \ref{sec:positioning_results}, 
with $43$ \glspl{AN}, user density of \SI{1000}{users/km}$^2$,
and all users are dropped with a uniform distribution on the simulated street area. To follow the 3GPP requirements and scenarios for next generation studies~\cite{3GPP_TS_38_913}, a single unpaired \SI{200}{MHz} \gls{TDD} carrier is assumed and \SI{30}{km/h} velocity is used for the \glspl{UE}. Additionally, the ratio between \gls{DL} and \gls{UL} is configured to \SI{4.7}:\SI{1} in the employed 5G \gls{TDD} frame structure~\cite{kela_location_based_beamforming_2016}. Every \gls{DL} transmission is assumed to start with a precoded \gls{DL} pilot and \gls{MRC} is used for {\colorrev calculating} the receive filter according to the measured \gls{DL} pilot. In case of location-based receive beamforming, estimated elevation and azimuth angles relative to \gls{UE}'s coordinate system are used for calculating the receive filter towards the serving \gls{AN}. For both location-based transmit and receive beamforming, a \SI{2}{} degree measurement error in addition to the \gls{UL} pilot measurement aging in both elevation and azimuth angles is assumed. \gls{UL} pilots used for \gls{CSIT} estimation and positioning are scheduled according to the round-robin scheme. Hence, in the simulated scenario the average \gls{CSIT} latency is $\sim$\SI{3.3}{ms}. \glspl{UE} are assigned to be served by the closest \gls{AN}, i.e., a centralized mobility management scheme based on estimated \gls{UE} locations is assumed.

\subsection{Performance Results and Comparison of Location-based and CSI-based Beamforming}

In~\cite{kela_location_based_beamforming_2016},
it was observed that both \gls{MF} and \gls{ZF} precoders work rather well in \glspl{UDN}, where \gls{LoS}-paths are dominating over reflections and diffractions. Hence, for this study a \gls{BD} algorithm~\cite{spencer_block_diagonalization}, which can be understood as an extension of \gls{ZF}, is chosen instead of conventional \gls{ZF}. Especially attractive feature in \gls{BD} is that the beams can be optimized for multiantenna receivers enabling better performance of receive filters. 

\begin{figure}[!t]
\centering
\includegraphics[width=\columnwidth]{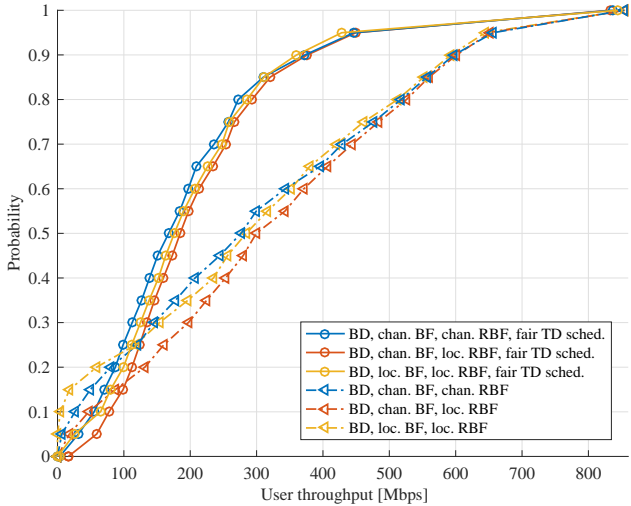}
\caption{\gls{DL} user throughput \glspl{CDF},  over random routes through the Madrid map, with channel/\gls{CSI} and location-based \gls{MU-MIMO} transmit \gls{BF} and \gls{RBF}. Relying only on location information is better on average than using only {\colorrev\gls{CSI}} measurements. The best overall performance is obtained by using channel-based \gls{BF} and location-based \gls{RBF}.}
\label{fig_dl_tput}
\vspace{-6pt}
\end{figure}

In Fig.~\ref{fig_dl_tput}, \glspl{CDF} of user experienced \gls{DL} throughputs are shown with both \gls{CSI}-based and location-based transmit and receive beamforming schemes. Due to high \gls{LoS} probability and dominance of \gls{LoS}-paths, both \gls{CSI}-based and location-based beamforming schemes obtain rather similar performance over the whole distribution. Additionally, focusing the receive filter only to \gls{LoS}-path with location-based receive beamforming outperforms \gls{DL} pilot based receive beamforming. Furthermore, ~$100$\% increase in 5-percentile throughput can be obtained when compared to the \gls{CSI}-based approach. Since \glspl{AN} are using same physical resources for transmitting beamformed \gls{DL} pilots, \gls{DL} pilot contamination degrades the performance of \gls{CSI}-based receive beamforming. In case of transmit beamforming with \gls{ZF}-based precoders like \gls{BD}, better performance at 5-percentile can be obtained with channel-based transmit beamforming due to the fact that there are still a few \glspl{UE} in \gls{NLoS} condition towards the serving \gls{AN}. Thus, the best overall performance is obtained by using channel information for transmit beamforming and location information for the receive filter. Note that the pilot overhead is here the same for all beamforming schemes. However, if pilot overhead caused by full-band reference signals needed in \gls{CSI}-based beamforming was reduced in the {\colorrev corresponding} location-based schemes, the performance would improve in terms of the mean throughput and area capacity as shown in~\cite{kela_location_based_beamforming_2016}. This is because location-based beamforming schemes do not require full-band reference signals, while narrowband, even single-subcarrier, pilots suffice.

In this example, in order to increase fairness of \gls{BD} precoding and to reach the throughput requirement of \SI{50}{Mbps} for all users all the time, {\colorrev stated in}~\cite{ngmn_5g}, the scheduling method introduced in \cite{kela_borderless_2015} is used. This fair \gls{TD} scheduling approach is applied in a way that in every other subframe only a subset of users is chosen as scheduling candidates, in particular the users with the lowest past average throughput. In other subframes the number of simultaneously served users is maximized to increase the total system throughput. 
{\colorrev The results in Fig.~\ref{fig_dl_tput} indicate that such scheduling provides less variation in throughputs across the \glspl{UE}.}
Moreover, the fair \gls{TD} scheduling with channel-based transmit beamforming and location-based receive beamforming decreases the simulated area throughput from~\SI{1}{Tbps/km\textsuperscript{2}} to~\SI{0.65}{Tbps/km\textsuperscript{2}}. {\colorrev Hence, when more fair \gls{TD} scheduling is applied, the total system throughput suffers to a certain extent from favoring users with poor channel conditions over users with high signal-to-interference-plus-noise ratio.} 

\section{Conclusions and Future Work} \label{sec:conclusion}

In this article, the prospects and enabling technologies for high-efficiency device positioning and location-aware communications in dense 5G networks were discussed and described. It was demonstrated that very high accuracy 2D/3D positioning and tracking can be accomplished, by adopting \gls{DoA} and \gls{ToA} estimation in the \glspl{AN} together with appropriate fusion filtering in the form of \gls{EKF}, for example. In general, outdoor positioning accuracies below one meter were shown to be technically feasible. It was also shown that location information can be used efficiently in the radio network, e.g., for geometric location-based beamforming, where the needed pilot or reference signal overhead is substantially smaller compared to the basic \gls{CSI}-based beamforming approaches. Thus, extracting and tracking the locations of the user devices in the 5G radio network can offer substantial benefits and opportunities for location-based services, in general, as well as to enhanced and more efficient communications and radio network management.

\ifCLASSOPTIONcaptionsoff
  \newpage
\fi



%
\bibliographystyle{IEEEtran/bibtex/IEEEtran}
\bibliography{IEEEtran/bibtex/IEEEabrv,main}




\end{document}